\documentclass{moriond}

\def\be{\begin{equation}}
\def\ee{\end{equation}}
\def\bea{\begin{eqnarray}}
\def\eea{\end{eqnarray}}

\newcommand{\Photo}{}

\begin{document}
\vspace*{4cm}
\title{Lepton Flavour Universality tests in $B$ decays as a probe for New Physics}

\author{C.\ Langenbruch on behalf of the LHCb collaboration}

\address{
  RWTH Aachen, I.\ Physikalisches Institut B, Sommerfeldstr. 14,\\
  52056 Aachen, Germany}

\maketitle\abstracts{
  In the Standard Model (SM), the coupling of the electroweak gauge bosons to the leptons is lepton flavour universal.
  Tests of this property constitute sensitive probes for new physics models that violate lepton flavour universality.
  Recent tests of lepton universality in rare $b\to s\ell\ell$ decays and semileptonic $b\to c\tau\bar{\nu}_\tau$ transitions have shown some tensions with the precise SM predictions. 
  These proceedings summarise the latest results on lepton flavour universality from the LHCb experiment. 
}

\section{Introduction}
Lepton flavour universality, \textit{i.\,e.}\ the equal coupling of the electroweak gauge bosons to the charged leptons, is a central property of the Standard Model (SM).
Lepton flavour universality is well established, for example in $\tau$ and $K$ decays~\cite{Ablikim:2014uzh,Lazzeroni:2012cx},
and deviations from the precise predictions would constitute unambiguous signs of New Physics (NP) beyond the SM.  

Measurements of rare $b\to s\ell\ell$ decays are particularly sensitive probes for NP effects as they are loop-suppressed in the SM and
effects beyond the SM could thus be comparably large. 
Furthermore, tree-level $b$-hadron decays involving $\tau$ leptons are currently under intense study, 
as many NP models predict enhanced couplings to the heavy third generation. 
In both cases it is useful to study ratios of branching fractions to final states containing different leptons, 
as hadronic uncertainties largely cancel. 

The LHCb experiment is ideally suited for lepton universality tests in $b$-hadron decays due to its large acceptance,
high trigger efficiencies and excellent tracking and particle identification capabilities. 
These proceedings give an overview of recent lepton universality tests at LHCb using the LHC run~1 data sample corresponding to an integrated luminosity of $3\,{\rm fb}^{-1}$. 

\section{Lepton flavour universality tests in rare decays}
\subsection{$R_{K^*}$}
The branching fraction ratio $R_{K^*}={\cal B}(B^0\to K^{*0}\mu^+\mu^-)/{\cal B}(B^0\to K^{*0}e^+ e^-)$ is a sensitive test of lepton universality. 
In the $q^2$ range $1.1<q^2<6\,{\rm GeV}^2$, where $q^2$ is defined as the squared invariant mass of the dilepton system $q^2=m^2(\ell\ell)$, 
$R_{K^*}$ is precisely predicted to be unity in the SM. 
Hadronic uncertainties largely cancel in the ratio and QED effects are not expected to exceed ${\cal O}(\%)$~\cite{Bordone:2016gaq}. 

\begin{figure}
  \begin{center}
    \includegraphics[height=4.25cm]{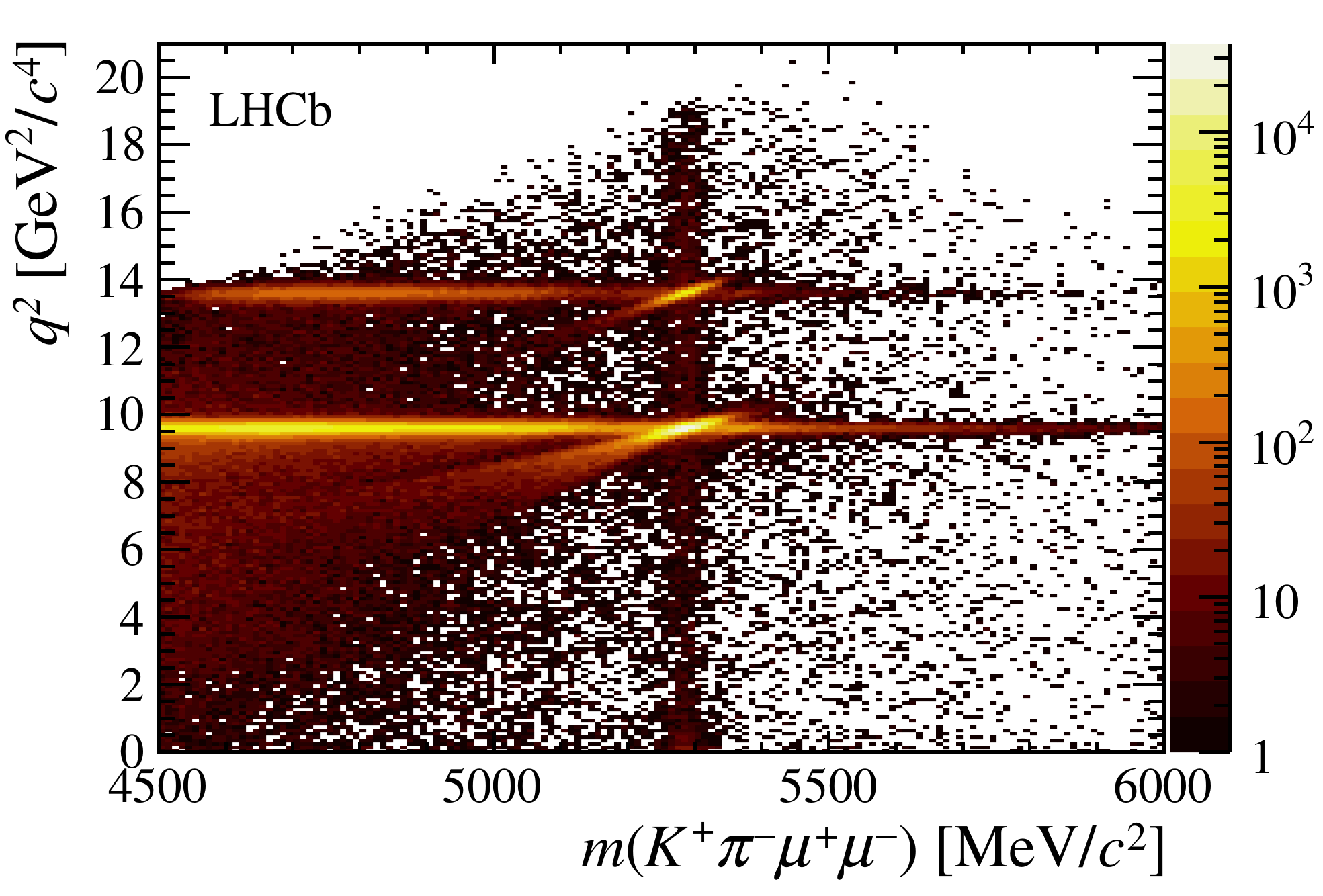}
    \includegraphics[height=4.25cm]{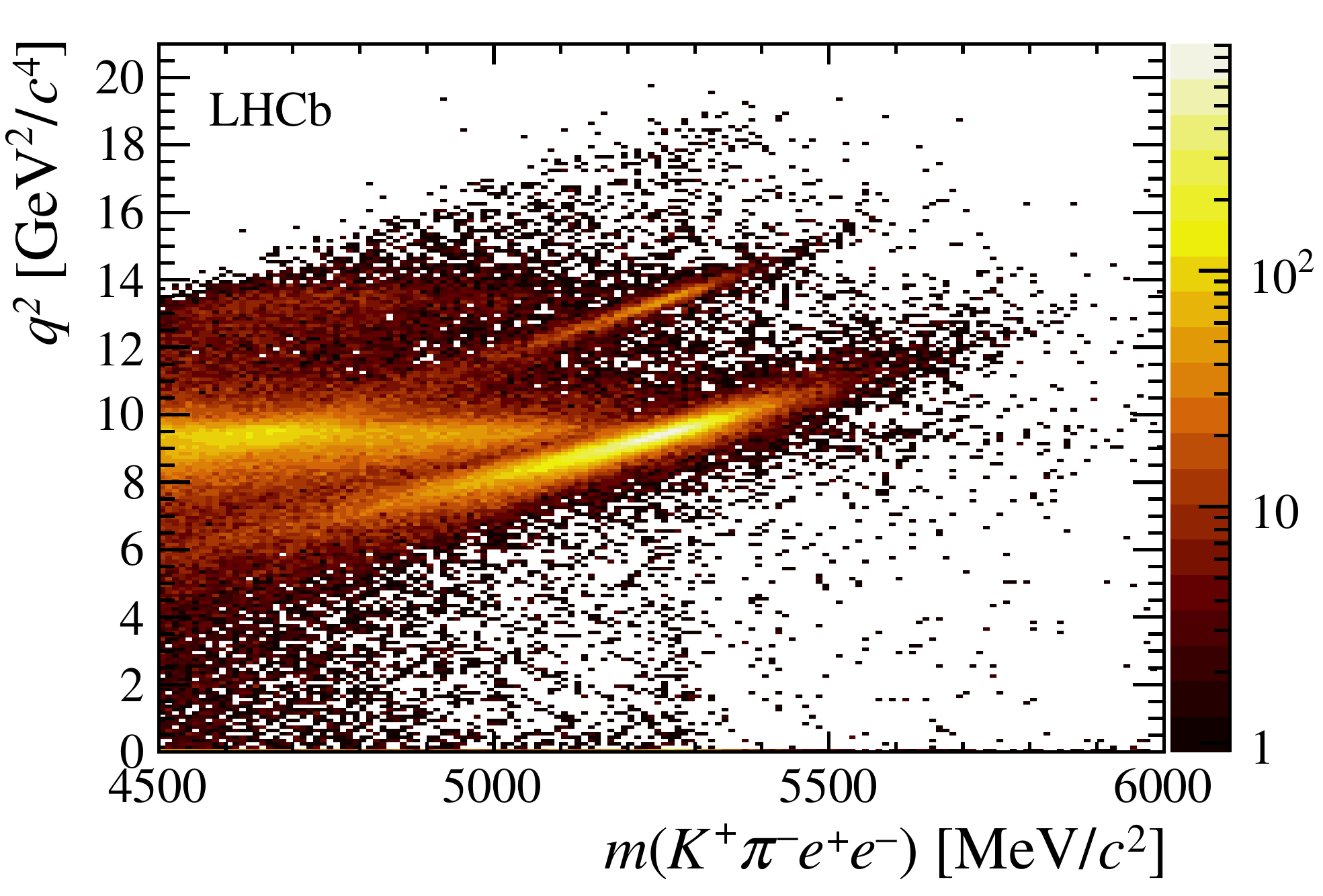}    
  \end{center}
  \caption{Reconstructed $K^{*0}\ell^+\ell^-$ mass vs.\ $q^2$ for (left) the decay $B^0\to K^{*0}\mu^+\mu^-$ and (right) the decay $B^0\to K^{*0}e^+e^-$~\protect\cite{Aaij:2017vbb}. 
    The control decays $B^0\to K^{*0} J/\psi(\to \ell^+\ell^-)$ are clearly visible, as is the more pronounced Bremsstrahlung for the electron final state.\label{fig:q2mb}}
\end{figure}

The LHCb measurement uses the double ratio with the tree-level decay $B^0\to K^{*0} J/\psi(\to \ell^+\ell^-)$ to determine $R_{K^*}$ according to 
\begin{eqnarray*}
  R_{K^*} &= \frac{{\cal B}(B^0\to K^{*0}\mu^+\mu^-)}{{\cal B}(B^0\to K^{*0}J/\psi(\to \mu^+\mu^-))} \bigg/
  \frac{{\cal B}(B^0\to K^{*0}e^+e^-)}{{\cal B}(B^0\to K^{*0}J/\psi(\to e+e^-))}
\end{eqnarray*}
This is experimentally advantageous, as many systematic uncertainties cancel in the double ratio.
Figure~\ref{fig:q2mb} shows the reconstructed $K^{*0}\ell^+\ell^-$ mass vs.\ $q^2$ for the decays $B^0\to K^{*0}\mu^+\mu^-$ and $B^0\to K^{*0}e^+e^-$. 
Experimentally, the electron mode $B^0\to K^{*0}e^+e^-$ is more challenging to reconstruct due to higher trigger thresholds
and more pronounced emission of Bremsstrahlung which deteriorates the mass resolution.

The tree-level decay $B^0\to K^{*0} J/\psi(\to \ell^+\ell^-)$ is used to control and correct simulation and to model the $B^0\to K^{*0}\ell^+\ell^-$ signal mass shape. 
Furthermore, it allows the determination of the single ratio $r_{J/\psi}={\cal B}(B^0\to K^{*0}J/\psi(\to \mu^+\mu^-))/{\cal B}(B^0\to K^{*0}J/\psi(\to e^+e^-))$.
This ratio is found to be compatible with unity and flat in control variables.
As systematic differences between electron and muon reconstruction do not cancel in this ratio, $r_{J/\psi}$ represents a very stringent crosscheck.  

\begin{figure}
  \begin{center}
    \includegraphics[height=4.25cm]{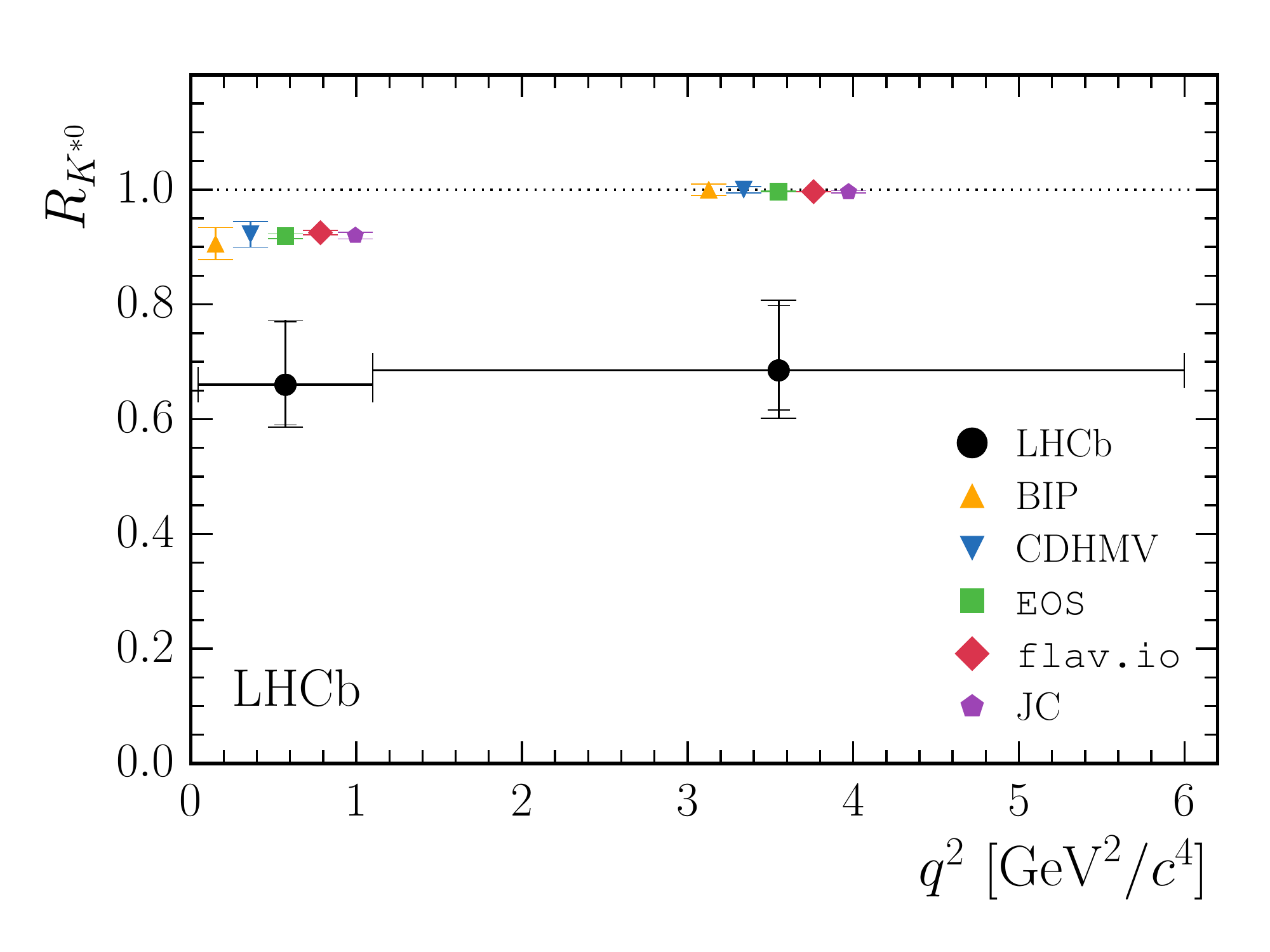}
    \includegraphics[height=4.25cm]{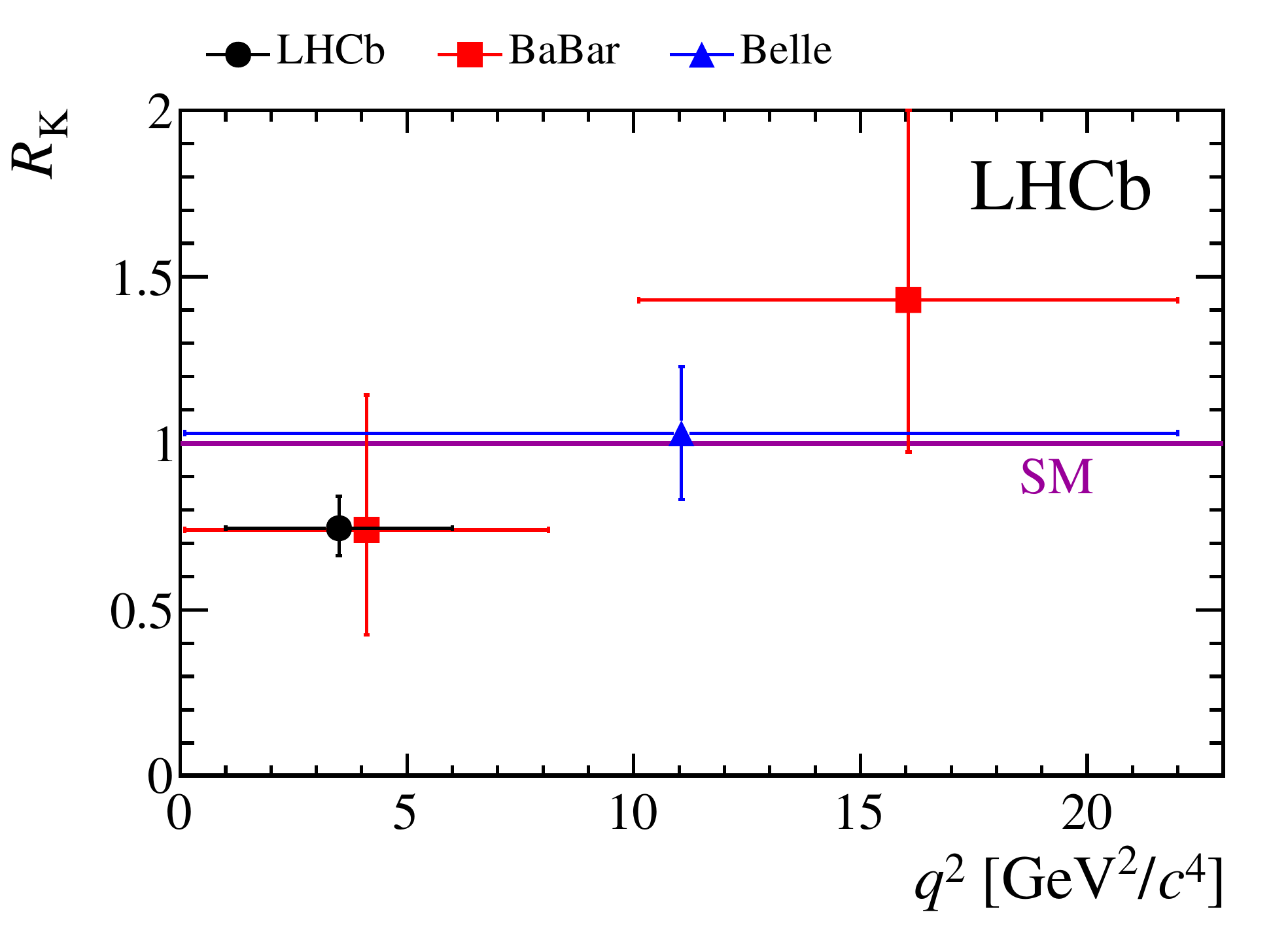}    
  \end{center}
  \caption{(Left) The LHCb measurement of the branching fraction ratio $R_{K^*}$ in two different $q^2$ bins~\protect\cite{Aaij:2017vbb}, overlayed with SM predictions~\protect\cite{Bordone:2016gaq,Capdevila:2016ivx,Serra:2016ivr,Altmannshofer:2017fio,Jager:2014rwa}. 
    (Right) Measurement of $R_K$ from LHCb~\protect\cite{Aaij:2014ora}, together with earlier measurements by the Belle and BaBar collaborations~\protect\cite{Wei:2009zv,Lees:2012tva}.\label{fig:rkstarrk}}
\end{figure}

LHCb measures $R_{K^*}$ in two distinct regions of $q^2$ and finds
\begin{eqnarray*}
R_{K^*}(0.045<q^2<1.1\,{\rm GeV^2}) &= 0.66^{+0.11}_{-0.07}\pm 0.03\\
R_{K^*}(1.1<q^2<6.0\,{\rm GeV^2}) &= 0.69^{+0.11}_{-0.07}\pm 0.05,
\end{eqnarray*}
in tension with the SM predictions at $2.1$--$2.3$ and $2.4$--$2.5\,\sigma$, respectively~\cite{Aaij:2017vbb}.  
The result is shown in Fig.~\ref{fig:rkstarrk}, overlaid with SM predictions~\cite{Bordone:2016gaq,Capdevila:2016ivx,Serra:2016ivr,Altmannshofer:2017fio,Jager:2014rwa}. 

\subsection{$R_K$}
The observable $R_K={\cal B}(B^+\to K^+\mu^+\mu^-)/{\cal B}(B^+\to K^+e^+e^-)$ is closely related to $R_{K^*}$.
As for $R_{K^*}$, hadronic uncertainties cancel in the ratio and it is thus a very clean test of lepton universality. 
LHCb finds $R_K=0.745^{+0.090}_{-0.075}\pm 0.036$ in the $q^2$ range $1<q^2<6\,{\rm GeV}^2$, 
in tension with the SM prediction of unity at $2.6\,\sigma$~\cite{Aaij:2014ora}. 
Figure~\ref{fig:rkstarrk} shows this result, together with previous results from the Belle and BaBar collaborations~\cite{Wei:2009zv,Lees:2012tva}. 

\subsection{Interpretation}
\begin{figure}
  \begin{center}
    \includegraphics[height=4.5cm,clip=true,trim=0mm 205mm 0mm 3mm]{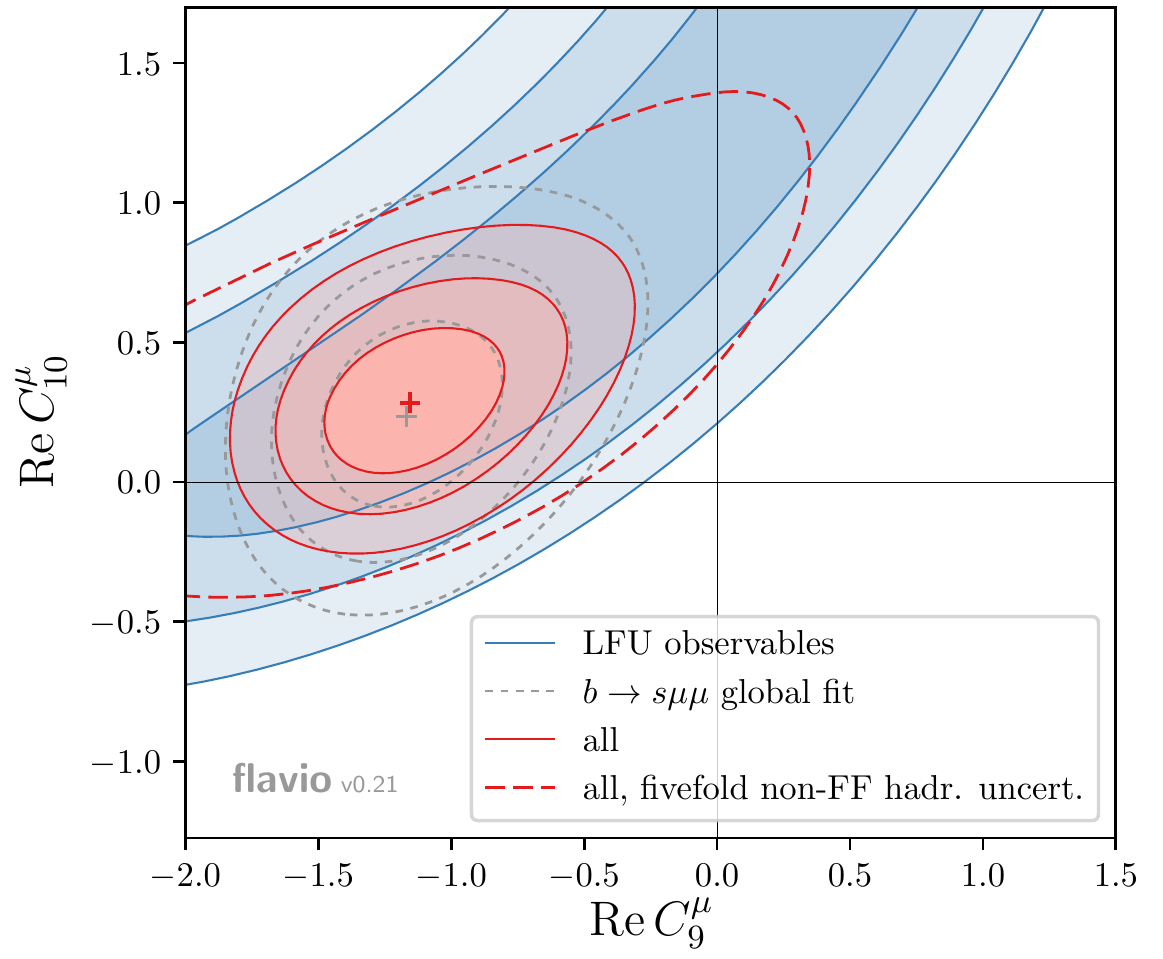}
    \hspace*{0.5cm}
    \includegraphics[height=4.5cm]{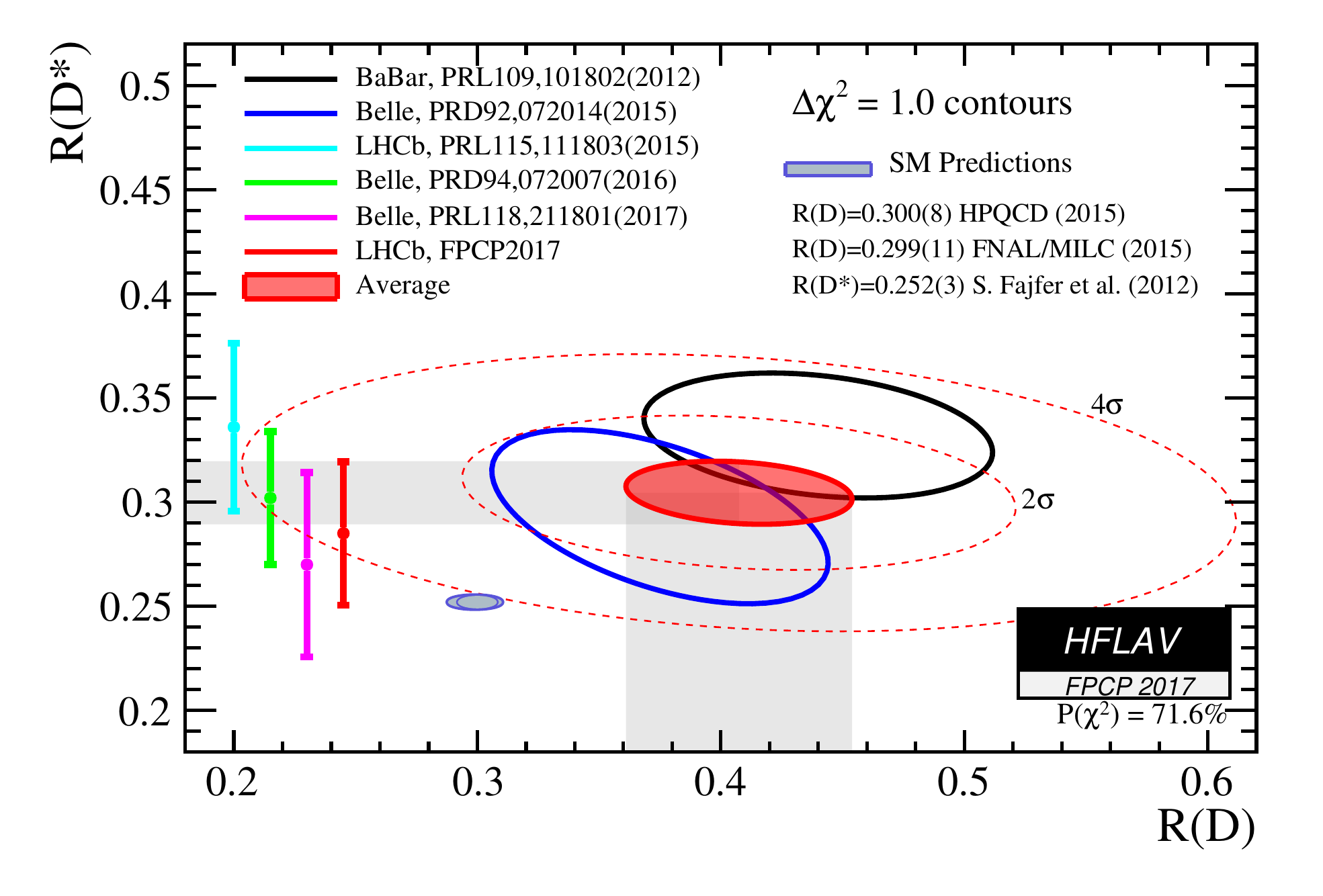}
  \end{center}
  \caption{(Left) Global fit of the effective vector and axial-vector couplings ${\cal C}_9$ and ${\cal C}_{10}$ using $b\to s\ell\ell$ data~\protect\cite{Altmannshofer:2017yso}.
    The blue contours only include lepton flavour universality tests~\protect\cite{Aaij:2017vbb,Aaij:2014ora,Wehle:2016yoi}, 
    the red contours also include further $b\to s\mu\mu$ observables~\protect\cite{Aaij:2014pli,Aaij:2015esa,Aaij:2016flj,Aaij:2015oid}.
    For other global fits see for example Refs.~\protect\cite{Capdevila:2017bsm,Hurth:2017hxg}.
    (Right) Combination by the Heavy Flavour Averaging Group~\protect\cite{Amhis:2016xyh} of the $R_{D^*}$ measurements from the LHCb collaboration~\protect\cite{Aaij:2015yra,Aaij:2017uff} and the $R_{D}$ and $R_{D^*}$ results from BaBar~\protect\cite{Lees:2012xj,Lees:2013uzd} and Belle~\protect\cite{Huschle:2015rga,Sato:2016svk,Hirose:2016wfn}.
    \label{fig:combinations}
  }
\end{figure}

The LHCb measurements of $R_K$ and $R_{K^*}$ are part of the so-called flavour anomalies in rare decays. 
The results can be interpreted in an effective field theory framework where they can be used to determine generalised effective couplings, the Wilson coefficients. 
Figure~\ref{fig:combinations} shows results for a combination of the lepton flavour universality tests $R_{K^{(*)}}$~\cite{Aaij:2017vbb,Aaij:2014ora} and a lepton universality test in $B^0\to K^{*0}\ell^+\ell^-$ angular observables by the Belle collaboration~\cite{Wehle:2016yoi}. 
The blue contours give the resulting confidence regions for the effective vector coupling ${\cal C}_9$ and the effective axial-vector coupling ${\cal C}_{10}$. 
The combination is in tension with the SM prediction at a level of around $4\,\sigma$~\cite{Altmannshofer:2017yso}. 

Adding additional $b\to s\mu\mu$ observables to the combination, 
including branching fraction measurements~\cite{Aaij:2014pli,Aaij:2015esa,Aaij:2016flj} and angular observables~\cite{Aaij:2015oid}, results in the red contours.
The significance of the tension with the SM further increases to above $5\,\sigma$,
it should however be noted that the theory uncertainties of some of the additional observables are currently under discussion. 

\section{Lepton flavour universality tests in tree-level decays}

\subsection{$R_{D^*}$ with leptonic $\tau$ reconstruction}
\begin{figure}
  \begin{center}
    \includegraphics[height=4.2cm,clip=true,trim=3mm 1mm 1mm 147.5mm]{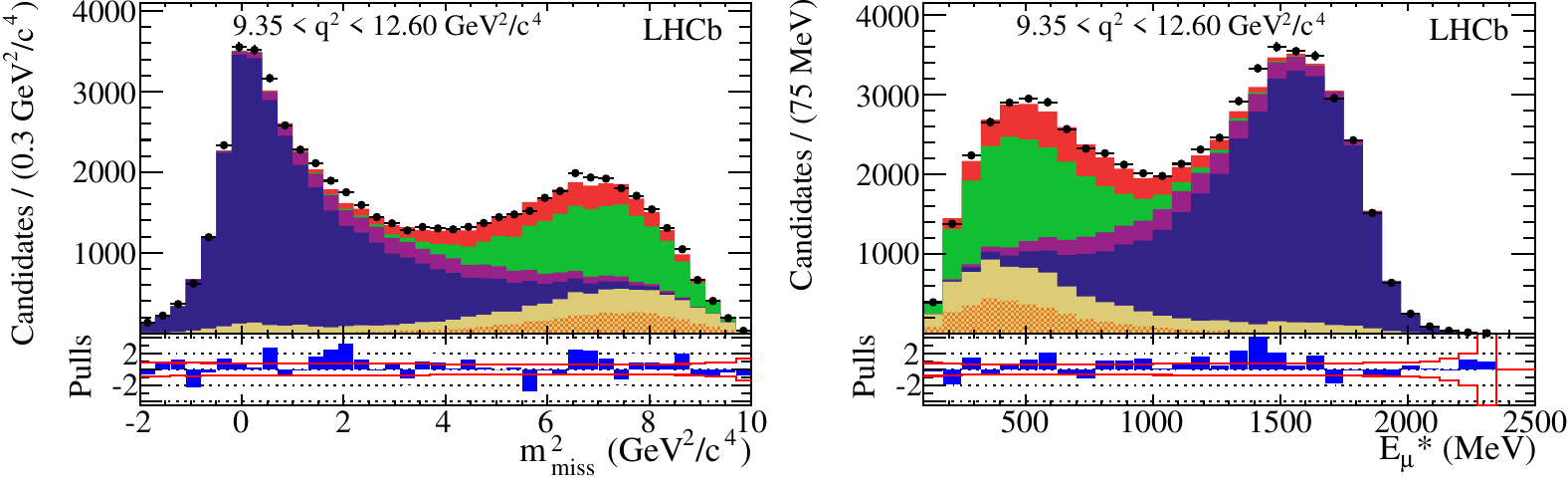}%
    \raisebox{2.7cm}{\includegraphics[height=1.5cm,clip=true,trim=38mm 164.5mm 91.2mm 7.7mmmm]{figs/rdst_leptonic.pdf}}
  \end{center}
  \caption{Distributions of $m_{\rm miss}^2$ and $E_\mu^*$ for the $q^2$ range $9.35<q^2<12.60\,{\rm GeV}^2$ for the LHCb measurement of $R_{D^*}$ using the leptonic $\tau$ decay mode~\protect\cite{Aaij:2015yra}.\label{fig:rdstleptonic}}
\end{figure}

The observable $R_{D^*}$ is defined as the ratio $R_{D^*}={\cal B}(\bar{B}^0\to D^{*+}\tau^-\bar{\nu}_\tau)/{\cal B}(\bar{B}^0\to D^{*+}\mu^-\bar{\nu}_\mu)$.
In the SM it can be precisely predicted to $R_{D^*}=0.252\pm 0.003$~\cite{Fajfer:2012vx}, it is lower than unity due to the considerable lepton mass difference. 
LHCb has measured $R_{D^*}$ in the leptonic $\tau$ decay mode $\tau^-\to \mu^-\bar{\nu}_\mu\nu_\tau$~\cite{Aaij:2015yra}. 
Due to the neutrinos in the final state, the $B$ momentum can not be reconstructed analytically.
Instead, vertexing information is used to deduce the $B$ flight direction which then allows to approximate the $B$ momentum with a resolution of around $18\%$.
The three quantities $m_{\rm miss}^2$, $E_\mu^*$ and $q^2$ are used in a three-dimensional template fit to separate the relative contributions of
signal mode $\bar{B}^0\to D^{*+}\tau^-\bar{\nu}_\tau$ and normalisation mode $\bar{B}^0\to D^{*+}\mu^-\bar{\nu}_\mu$. 
Here, $m_{\rm miss}^2$ denotes the missing mass squared, $E_\mu^*$ the muon energy in the $B$ rest frame and $q^2$ the four-momentum transfer squared.
Background and signal shapes are extracted from control samples and simulations are validated against data. 
Figure~\ref{fig:rdstleptonic} shows $m_{\rm miss}^2$ and $E_\mu^*$ for the highest $q^2$ bin.
The  resulting value is $R_{D^*}=0.336\pm 0.027 \pm 0.030$ which is compatible with the SM prediction at $2.1\,\sigma$. 

\subsection{$R_{D^*}$ with hadronic $\tau$ reconstruction}
\begin{figure}
  \begin{center}
    \includegraphics[height=4.25cm,clip=true,trim=10mm 5mm 5mm 185.3mm]{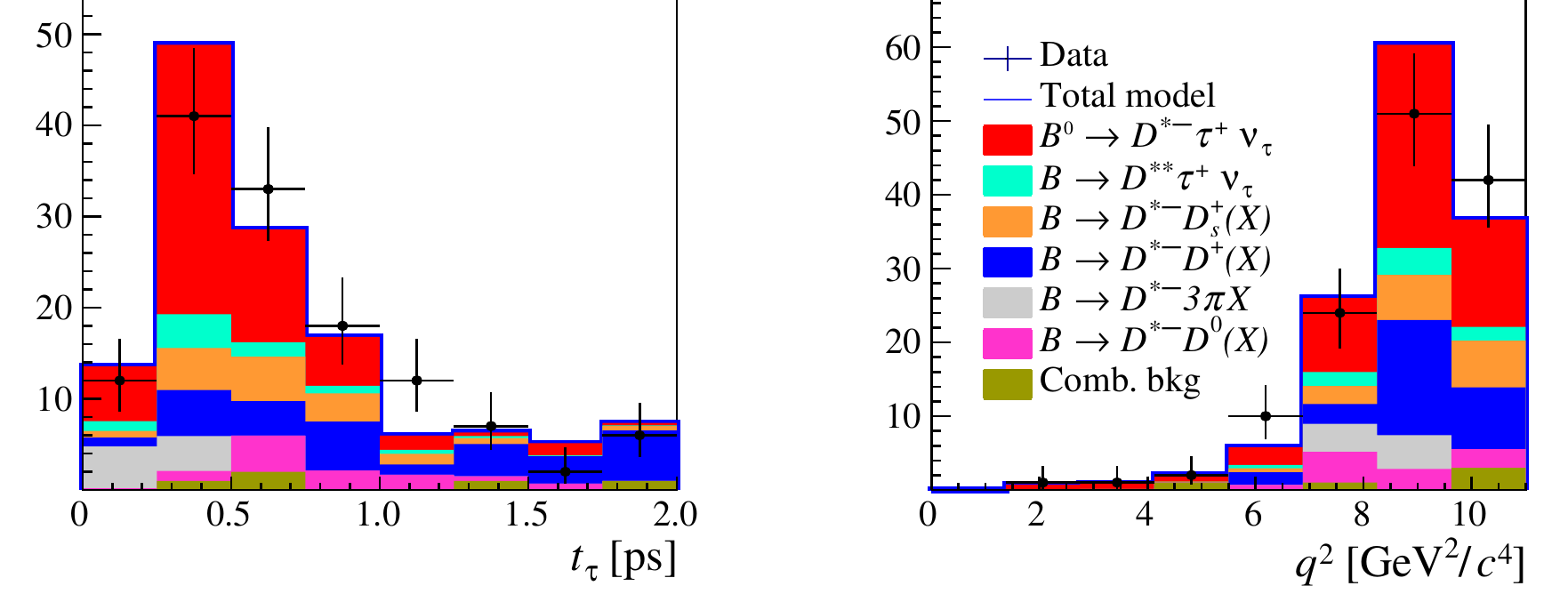}
  \end{center}
  \caption{(Left) The $\tau$ lifetime and (right) $q^2$ in the bin of the multivariate classifier with the highest purity for the LHCb measurement of $R_{D^*}$ using the hadronic $\tau$ decay mode~\protect\cite{Aaij:2017uff}.\label{fig:rdsthadronic}}
\end{figure}

A more recent determination of $R_{D^*}$ at LHCb uses the hadronic $\tau^-\to \pi^+\pi^-\pi^-(\pi^0)\nu_\tau$ decay mode~\cite{Aaij:2017uff}. 
In this case, the well known decay $B^0\to D^{*-}3\pi$ is used as normalisation mode which has the same final state as the signal. 
Experimentally, this is advantageous as many systematic uncertainties cancel.
Backgrounds from hadronic $B\to D^{*}3\pi X$ decays are suppressed by exploiting the significant $\tau$ lifetime,
and multivariate techniques are used to suppressed doubly-charmed $B\to D^{*-}D_{(s)}X$ background. 
A three-dimensional template fit in $q^2$, the $\tau$ lifetime and the output of a multivariate classifier is performed. 
The templates are derived from simulation and validated on control samples.
Figure~\ref{fig:rdsthadronic} shows the $\tau$ lifetime and $q^2$ in the bin of the multivariate classifier with the highest signal purity. 
As external inputs the branching fractions of the normalisation mode ${\cal B}(B^0\to D^{*-}3\pi)=(7.21\pm 0.28)\times 10^{-3}$
as well as ${\cal B}(B^0\to D^{*-}\mu^+\nu_\mu)=(4.88\pm 0.10)\times 10^{-2}$ are used~\cite{Aaij:2013xca,TheBABAR:2016vzj,Majumder:2004su,Amhis:2016xyh}. 
The resulting value 
is $R_{D^*}=0.291\pm 0.019 \pm 0.026 \pm 0.013$, 
where the last uncertainty is due to the external inputs.  
The result is $1\,\sigma$ above the SM prediction. 

\subsection{$R_{J/\psi}$}
Due to the different hadron species produced at the LHC, LHCb is also able to perform lepton universality tests with $B_c^+$ mesons. 
LHCb has performed a measurement of the ratio $R_{J/\psi}={\cal B}(B_c^+\to J/\psi \tau^+ \nu_\tau)/{\cal B}(B_c^+\to J/\psi \mu^+ \nu_\mu)$,
where the $\tau^+$ is reconstructed in the leptonic decay $\tau^+\to\mu^+\nu_\mu\bar{\nu}_\tau$~\cite{Aaij:2017tyk}. 
The predictions for this observable are less precise than for $R_{D}$ and $R_{D^*}$ due to the larger form-factor uncertainties~\cite{Anisimov:1998uk,Kiselev:2002vz,Ivanov:2006ni,Hernandez:2006gt}, with central values in the range $R_{J/\psi}\in [0.25,0.28]$. 
A three-dimensional template fit is performed to the quantities $m_{\rm miss}^2$, $\tau(B_c^+)$ and $Z=(E_\mu^*,q^2)$. 
The resulting value of $R_{J/\psi}=0.71\pm 0.17 \pm 0.18$ is around $2\,\sigma$ above the SM prediction.
This analysis also constitutes the first evidence for the decay $B_c^+\to J/\psi\tau^+\nu_\tau$ at $3\,\sigma$. 

\subsection{Combination}
The measurements of $R_{D^*}$ by LHCb~\cite{Aaij:2015yra,Aaij:2017uff} can be combined with measurements of $R_D$ and $R_{D^*}$ by the BaBar~\cite{Lees:2012xj,Lees:2013uzd} and Belle~\cite{Huschle:2015rga,Sato:2016svk,Hirose:2016wfn} collaborations. 
Figure~\ref{fig:combinations} shows the resulting confidence regions, derived by the Heavy Flavour Averaging Group~\cite{Amhis:2016xyh}. 
The $R_D$ and $R_{D^*}$ combination, given by the red contours, 
is in tension with the SM prediction given by the blue contours at $4.1\,\sigma$. 
It should be noted that the tension is somewhat reduced using recent theory input~\cite{Bigi:2017jbd}. 

\section{Conclusions and outlook}
Recent results on lepton universality in $B$ decays show intriguing tensions with the precise SM predictions. 
In rare $b\to s\ell\ell$ decays a combined tension of around $4\,\sigma$ is found for the ratios $R_K$ and $R_{K^*}$ measured by the LHCb collaboration. 
Consistent explanations of the anomalies in the rare decays are possible, 
examples are new heavy gauge bosons~\cite{Altmannshofer:2014cfa,Crivellin:2015mga,Celis:2015ara,Falkowski:2015zwa} and leptoquarks~\cite{Hiller:2014yaa,Gripaios:2014tna,Varzielas:2015iva,Barbieri:2016las}. 
In semileptonic $b\to c\tau\bar{\nu}_\tau$ transitions, multiple experiments find tensions with the SM. 
A combination of the results from the LHCb, BaBar and Belle collaborations yields a significance of $4.1\,\sigma$. 
Combined explanations of the flavour anomalies in rare and tree-level decays are challenging but possible~\cite{Bhattacharya:2014wla,Fajfer:2015ycq,Greljo:2015mma,Barbieri:2015yvd,Bauer:2015knc,Boucenna:2016qad,Becirevic:2016yqi,Crivellin:2017zlb}. 

Since the SM predictions for the presented lepton flavour universality tests are generally very precise,
the currently limiting factor are the experimental uncertainties. 
Future analyses of the LHCb experiment exploiting the data samples collected during the LHC Run~2 and the LHCb upgrade will reduce the experimental uncertainties for the lepton universality tests in both rare loop-level as well as tree-level decays.
Furthermore, LHCb will test lepton universality in additional decay modes (e.g.\ measurements of $R_\phi$, $R_{K\pi\pi}$ and  $R_{\Lambda^{(*)}}$ in the rare decays and of $R_{D^0}$, $R_{D^+}$, $R_{D_s}$ and $R_{\Lambda_c^{(*)}}$ in tree-level decays), 
and important related searches for lepton flavour violating decays will be performed. 
Independent clarification of the anomalies is also expected from future lepton universality measurements by the Belle~II experiment. 

\section*{Acknowledgments}
C.\,L.\ gratefully acknowledges support by the Emmy Noether programme of the Deutsche Forschungsgemeinschaft (DFG), grant identifier LA 3937/1-1.

\end{document}